\documentclass[prl,aps,superscriptaddress,twocolumn,notitlepage,showpacs]{revtex4-2}

\usepackage{graphicx}
\usepackage{dcolumn}
\usepackage{bm}
\usepackage{svg}
\usepackage{amsmath}
\usepackage{graphicx}
\usepackage{caption}
\usepackage{subfigure}
\usepackage{float}
\usepackage{mathrsfs}
\usepackage{color}
\usepackage[justification=centering,
            format=plain]{caption}
\usepackage{latexsym}

\begin{document}


    \title{Entangled Photons Enabled Ultrafast Stimulated Raman Spectroscopy for Molecular Dynamics}

\author{Joel Jiahao Fan}
\affiliation{Department of Physics, City University of Hong Kong, Kowloon, Hong Kong SAR}

\author{Zhe-Yu Jeff Ou}
\email{jeffou@cityu.edu.hk}
\affiliation{Department of Physics, City University of Hong Kong, Kowloon, Hong Kong SAR}

\author{Zhedong Zhang}
\email{zzhan26@cityu.edu.hk}
\affiliation{Department of Physics, City University of Hong Kong, Kowloon, Hong Kong SAR}
\affiliation{City University of Hong Kong, Kowloon, Shenzhen Reseach Institute, Shenzhen, Guangdong 518057, China}

\date{\today}

\begin{abstract}
Quantum entanglement has emerged as a great resource for interactions between molecules and radiation. 
We propose a new paradigm of stimulated Raman scattering with entangled photons. A quantum ultrafast Raman spectroscopy is developed for condensed-phase molecules, to monitor the exciton populations and coherences. Analytic results are obtained, showing a time-frequency scale not attainable by classical light. The Raman signal presents an unprecedented selectivity of molecular correlation functions, as a result of the Hong-Ou-Mandel interference. This is a typical quantum nature, advancing the spectroscopy for clarity. Our work suggests a new scheme of optical signals and spectroscopy, with potential to unveil advanced information about complex materials.


\end{abstract}

\maketitle

{\it Introduction}.--With the advancements of quantum light sources, the study of spectroscopy and sensing draw much attention in a diversity of active research fields \cite{PhysRevLett.93.023005,PhysRevA.69.013806,PhysRevLett.94.043602,doi:10.1063/1.5138691,Lee2006EntangledPA,upton2013optically,harpham2009thiophene,guzman2010spatial,upton2013optically}. Quantum states of light with different types of entanglement offer new freedom for the light-matter interactions, spectroscopy and precise measurement \cite{Ou:87,PhysRevLett.62.1603,PhysRevA.54.R4649,PhysRevA.50.67,Mitchell_2004,MUKAMEL2011132,Schlawin_2017}. Novel knobs can be developed thereby for controlling the atom and molecule motions at microscopic scale. A capability of controlling the multi-photon transitions, as a signature of nonlinear optical processes, was enabled by photon entanglement \cite{PhysRevLett.80.3483,PhysRevLett.93.093002,PhysRevLett.125.133601}. Much attention has been drawn recently for the importance of entangled photons in various fields including quantum simulations \cite{doi:10.1126/science.aab0097} and further the marriage of molecular spectroscopy with quantum photonics.

The multi-photon interactions with complex molecules were studied recently in a context of quantum-light spectroscopy. Several experiments indicated the extraordinary transitions with entangled two-photon absorption (ETPA)--the inhomogeneous line broadening can be circumvented for an efficient population of highly-excited states of molecules \cite{PhysRevLett.78.1679,doi:10.1021/ja1016816,PhysRevLett.80.3483,PhysRevLett.93.023005,doi:10.1021/jp066767g,doi:10.1021/acs.jpca.7b06450,PhysRevA.76.043813,doi:10.1021/ja803268s,doi:10.1021/acs.jpca.8b06312,PhysRevLett.123.023601, doi:10.1063/5.0049338,doi:10.1021/jacs.1c02514,doi:10.1021/acs.accounts.1c00687,doi:10.1021/acsphotonics.2c00255,doi:10.1063/5.0128249}. The multi-photon interaction has been studied much in atoms, it is however an open issue in molecules so far. This arises predominately from the couplings of electrons to more degrees of freedom, which bring up new challenges for the optical response. Elaborate experiments demonstrated the incredible power of entangled photons yielding the probe and control of electronic structures with unprecedented scales \cite{10.1063/5.0010909}. Recent studies extended the ETPA to time-resolved regime, showing miraculous cancellation of molecular correlation functions not accessible by classical pulses \cite{doi:10.1021/acsphotonics.1c01755}. New control knobs by entangled photons may enable considerable suppression of background in the radiation. An improvement in signal-to-noise ratio can be thus expected. The entanglement-refined interactions with molecules may induce a nonlinearity prominently resonant with the excited-state relaxation as well as the many-particle couplings. These call for a thorough understanding of the quantum-light interactions with complex molecules at ultrafast timescales.

The Raman process, as a typical component of the multi-photon interactions, closely connects to the quantum-light fields \cite{PhysRevA.77.022110,PhysRevA.79.033832}. Coherent Raman spectroscopy including a variety of schemes provides a powerful tool for quantum physics and molecular characterization. Extensive studies have demonstrated the quantum advantage of entangled light in spectroscopy \cite{MUNKHBAATAR2017581,doi:10.1063/5.0015432,https://doi.org/10.48550/arxiv.2212.11519}. The time-frequency entanglement of photons may enable a superresolved capability, free of the conjugation of temporal and spectral scales that posses the fundamental limit in Raman spectroscopy using classical light. A femtosecond CARS was proposed recently with entangled photon pairs, to monitor the ultrafast dyanamics of electronic coherence and the passage of conical intersections \cite{Zhang2022}. Furthermore, recent progress reported a CARS with squeezed photons in a nonlinear interferometer \cite{Michael_2019}. A quantum-enhanced measurement beyond the shot-noise limit was therefore performed. As a different scheme, the stimulated Raman scattering is sensitive to the molecular populations that are of fundamental interest and importance for the cooperative effects and multi-exciton correlations. These can be monitored in a greater way by making use of entangled photons and nonlinear interferometry \cite{doi:10.1021/jz501124a,PhysRevResearch.3.043029}. 

\begin{figure*}[t]
 \captionsetup{justification=raggedright,singlelinecheck=false}
\centering
\includegraphics[width=0.95\textwidth,height=6cm]{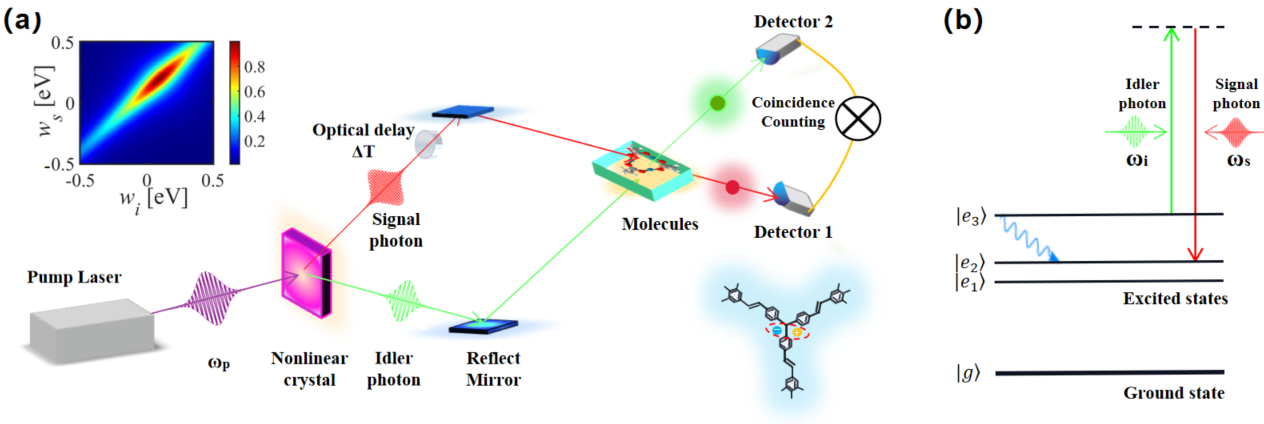}
\caption{(a) Schematic of entangled twin photons (correlated with each other rather than the anti-correlation) as an ultrafast probe for molecules, where the nonlinear mediums and the photon-coincidence counting measurement are presented; Small panel plots $|\Phi(\omega_s,\omega_i)|$ of the entangled twin photons. (b) Level scheme of molecular relaxation interacting with two entangled photons that induce the stimulated Raman  scattering.}
\label{set up}
\end{figure*}

In this Letter, we propose an ultrafast stimulated Raman spectroscopy (USRS) using entangled photons. A microscopic theory is developed with molecular trimers. Here the molecules play an active role in interacting with optical signals, rather than the passive role of the beam splitters in quantum optics. Due to the entanglement, such a quantum USRS (Q-USRS) enables a superresolved nature of the spectrum with time-frequency scales not attainable by classical pulses. Moreover, as a result of the multi-photon quantum interference, the spectroscopic signals are presented. An unprecedented selectivity is thus shown by the signals, enabling a selective access to molecular correlation functions. This is a hard task for classical pulses, but useful in spectroscopy. Our work provides a new paradigm for Raman spectroscopy and metrology, insightful for the study of heterostructural materials.

{\it Quantum stimulated Raman scattering}.--We consider a generic model of molecules interacting with entangled photons generated by nonlinear mediums in Fig.\ref{set up}(a). The two entangled photons are shaped in a short pulse. The photons in s and i arms are jointly scattered by molecular excited states, inducing the stimulated Raman process. A coincidence counting of emission is measured, where no spectrometers are required.

The Raman interaction between molecules and entangled photons is of the form
 \begin{equation}
    V(t) = \alpha(t) E_s (t) E_i^{\dagger} (t) + \text{h.c.}
\label{Vt}
\end{equation}
where $E_s(t)$ and $E_i(t)$ play the roles of the respective pump and probe fields containing multiple frequency modes. $\alpha(t) = \sum_{m>n} \alpha_{mn} |\psi_m\rangle \langle \psi_n|(t) + \text{h.c.}$ defines the Raman polarizability operator and the elements $\alpha_{mn}$ are given in Supplemental Material (SM) \cite{SMM}. Usually $|\psi_m\rangle \langle \psi_n|(t) = |\psi_m\rangle \langle \psi_n| e^{i\omega_{mn}t}$ for closed systems but we will not adopt this assumption, so as to involve more general cases described by a reduced density matrix.

Eq.(\ref{Vt}) resembles the beam-splitter interaction, indicating the two-photon interference that essentially interplays with the molecular excitations. Further results will elaborate the active role of the Hong-Ou-Mandel (HOM) effect in Raman spectroscopy for a monitoring of excited-state dynamics. 

The Q-USRS is defined as the coincidence counting of the transmissions along s and i arms, and is given by 
$S(T_s,T_i) = \int dt \frac{d}{dt} \langle E_s^{\dagger}(t) E_i^{\dagger}(t) E_i(t) E_s(t) \rangle$, i.e., 
\begin{equation}
    \begin{split}
        & S(T_s,T_i) = \Re \iiiint_{-\infty}^{+\infty} d\omega' d\omega dt d\tau\ \theta(t-\tau) \\[0.15cm]
        & \qquad\quad \times \Big[ \langle \psi(\tau)|\alpha(t-\tau) \alpha|\psi(\tau)\rangle C_{\text{I}}(t,\tau; T_s,T_i) \\[0.15cm]
        & \qquad\quad \ \  + \langle \psi(\tau)|\alpha \alpha(t-\tau)|\psi(\tau)\rangle C_{\text{II}}(t,\tau; T_s,T_i)\Big]
    \end{split}
\label{SS}
\end{equation}
where $\psi(\tau)$ is the molecular wave function including full degrees of freedom. $T_s$ and $T_i$ are the central times of the pulses in s and i photons, respectively. $\Delta T = T_i - T_s$ measures the time delay between s and i photons that is controllable via optical paths. This provides the arrival times of the two photons with delays relative to the resonant pump that creates the electronic excitations in molecules, shown in Fig.1(b). 
The six-point field correlation functions are $C_{\text{I}}(t,\tau; T_s,T_i) = e^{i\omega(t-T_s)} {\cal F}_{\text{I}}(t,\tau; T_s,T_i) + (s\leftrightarrow i)$, $C_{\text{II}}(t,\tau; T_s,T_i) = e^{i\omega(t-T_s)} {\cal F}_{\text{II}}(t,\tau; T_s,T_i) + (s\leftrightarrow i)$ where ${\cal F}_{\text{I}} = \langle \Psi|{\cal N}_s(\omega'){\cal E}_i^{\dagger}(\omega)E_s(t-T_s)E_s^{\dagger}(\tau-T_s) E_i(\tau-T_i)|\Psi\rangle$,  ${\cal F}_{\text{II}} = \langle \Psi|E_s^{\dagger}(\tau-T_s) E_i(\tau-T_i){\cal N}_s(\omega'){\cal E}_i^{\dagger}(\omega)E_s(t-T_s)|\Psi\rangle$, and ${\cal N}_s(\omega') = {\cal E}_s^{\dagger}(\omega'){\cal E}_s(\omega')$.

\begin{figure}[t]
 \captionsetup{justification=raggedright,singlelinecheck=false}
\centering
\includegraphics[width=0.45\textwidth]{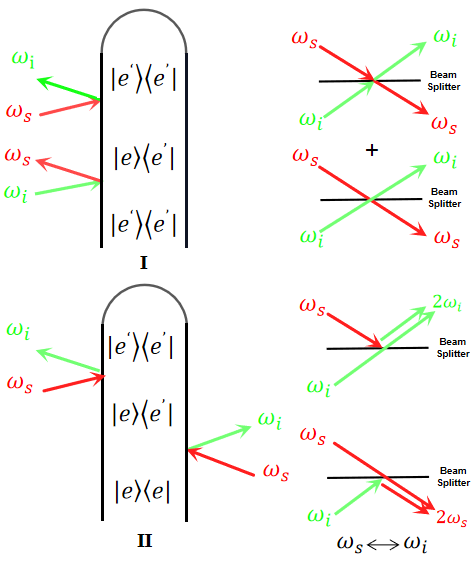}
\caption{(Top) Feynman's loop diagrams for the stimulated Raman scattering with twin photons; (I) parametric process and (II) dissipative process. (Bottom) Two pathways of the two-photon interference in a fashion of the Hong-Ou-Mandel (HOM) scheme corresponding to the processes (I) and (II). Notice that the complicated light-molecule interactions lead to the two-photon interference not exactly the same as the original HOM using 50:50 beam splitters.}
\label{energy level}
\end{figure}

The two components with $C_{\text{I}}$ and $C_{\text{II}}$ in Eq.(\ref{SS}) correspond to the loop diagrams in Fig.\ref{energy level} which govern the multipoint Green's functions of Raman operators. It includes the components: pathway I for parametric process and pathway II for dissipative process \cite{book}.

Notably from Eq.(\ref{SS}), $C_{\text{I}} \neq C_{\text{II}}$ in general for quantum fields whereas $C_{\text{I}}=C_{\text{II}}$ for classical fields. One can probably achieve the selectivity of the molecular correlation functions, not attainable by classical pulses.

The density matrix $\rho(\tau)=|\psi(\tau)\rangle\langle \psi(\tau)|$ contains full information about the structures and dynamics of molecules. These are essentially imprinted into the Raman signal, so as to be read out by making use of the quantum-field correlations.


{\it Q-USRS with entangled photons}.--The entangled state of photons in Fig.\ref{set up}(a) is of the form $|\Psi\rangle=\iint_{-\infty}^{+\infty} \text{d}\omega_s \text{d}\omega_i \Phi(\omega_s,\omega_i) a_{\omega_s}^{\dagger} a_{\omega_i}^{\dagger}|0\rangle$ and
\begin{equation}
    \Phi(\omega_s,\omega_i) = A(\omega_s - \omega_i - \omega_-)\phi\left[\frac{ k(\omega_s,\omega_i)L}{2}\right] e^{i k(\omega_s,\omega_i)L/2}
\label{Phi}
\end{equation}
is the two-photon wave function with a phase matching $k(\omega_s,\omega_i)L = \left(\omega_s - \frac{\omega_+}{2}\right)\tau_s + \left(\omega_i - \frac{\omega_+}{2}\right)\tau_i$ where $\tau_s (\tau_i)$ is the time delay of s(i)-arm photons relative to the pump field $A$, due to the group velocity dispersion in the nonlinear mediums. $A$ is a classical field with an effectively narrow bandwidth $\sigma_0$, so that $A(\omega_s - \omega_i - \omega_-)\rightarrow \delta(\omega_s - \omega_i - \omega_-)$ as $\sigma_0\rightarrow 0$. $\omega_s-\omega_i$ indicates quantum correlated photon pairs rather than the anti-correlated nature directly from the down-conversion process.


Notably, the entangled photon states cause a miraculous cancellation of the field correlation functions. In particular, $C_{\text{I}}\neq 0$ but $C_{\text{II}}=0$. This is a typical two-photon interference, arising from the HOM effect \cite{PhysRevLett.59.2044}. As a result, the parametric component in Fig.\ref{energy level} survives whereby the diagram II vanishes. A great selectivity of molecular Green's functions is thus expected, not accessible by the Raman spectroscopy using classical pulses. It should be aware of that the Raman interaction in Eq.(\ref{Vt}) largely differs from the 50:50 beam splitter in quantum optics. The destructive two-photon interference is therefore retained, giving $C_{\text{I}}\neq 0$ rather than causing the full cancellation as in the original HOM scheme. The constructive two-photon interference groups two photons in one output mode, making $C_{\text{II}}=0$. These yield an aspect of conceptual importance for quantum-light spectroscopy, because the residue part characterizes the spectral lines that may provide a key monitoring of molecular structures and relaxation. Elaborate simulations of the Q-USRS incorporating the two-photon interference will be performed later on.

On substituting Eq.(\ref{Phi}) into Eq.(\ref{SS}), we obtain the entangled Q-USRS, i.e.,
\begin{subequations}
    \begin{align}
        S & (\omega_-;T_s,T_i)  \propto \sum_{e,e'}\sum_{e''}  \iint dt d\tau e^{-i\omega_{e''e'}(t-\tau)} \rho_{ee'}(\tau) \nonumber \\[0.15cm]
        & \qquad\qquad \times \Big[\tilde{\Phi}^*(t-T_i,\tau-T_i) \tilde{\Phi}(\tau-T_s,t-T_i) \nonumber \\[0.15cm]
        & \qquad\qquad + \tilde{\Phi}^*(\tau-T_s,t-T_s) \tilde{\Phi}(t-T_s,\tau-T_i)\Big] \label{SR}
    \end{align}
\end{subequations}
where $\tilde{\Phi}(t_1,t_2) = \frac{1}{4\pi^2} \iint_{-\infty}^{+\infty} \Phi(\omega_1, \omega_2) e^{-i(\omega_1 t_1 +\omega_2 t_2)} \text{d} \omega_1 \text{d}\omega_2$ is the wave packet of the twin photons. The background has been dropped from Eq.(\ref{SR}), in that no spectral lines are produced.

Eq.(\ref{SR}) indicates the role of the quantum-entangled photons whose unusual band properties may provide versatile tool for controlling the fast excited-state dynamics of molecules.

{\it Superresolved nature}.--We will proceed from Eq.(\ref{SR}) in the regime of narrow-band field $A$, for some generic properties of the Q-USRS, e.g., superresolved time-frequency scales. Defining $\tilde{\phi}(t)=\frac{1}{2\pi} \int_{-\infty}^{+\infty} e^{-iv t}\phi(v) dv$ and assuming identical group velocity for s and idler photons, i.e., $\tau_s=\tau_i=\tau_0$, the two-photon wavepacket is $\tilde{\Phi}(t,t') = c(t,t') \tilde{\phi}[(t+t'-\tau_0)/\tau_0]$ and $c(t,t')$ is an oscillatory function of $t,t'$ \cite{SMM}. 
The twin photons thus present a pulse duration of $\tau_0$. Invoking the {\it impulsive approximation} such that $\tau_0$ is much shorter than the dephasing and solvent timescales, the Q-USRS reads $S (\omega_-;T_s,T_i) = {\cal S}(\omega_-;T_s,T_i) + (s\leftrightarrow i)$ with ($\omega_- := \omega_- + i\sigma_0$, $\omega_{e''e'} := \omega_{e''e'} - i\gamma_{e''e'}$)
\begin{equation}
    \begin{split}
        {\cal S}(\omega_-;T_s,T_i) \approx  \frac{e^{-\frac{i}{2}\varpi \Delta T}}{8 i \pi^2 \tau_0} \sum_{e,e'} \sum_{e''}
        \frac{\rho_{ee'}\left(T_i + \frac{\tau_0}{2}\right) W}{\omega_- - \omega_{e''e'} + \frac{\omega_{ee'}}{2}}
    \end{split}
\label{SRI}
\end{equation}
where $\varpi = \omega_-+\omega_+$ and $W$ is an overlap integral, i.e.,
\begin{equation}
    \begin{split}
        W = \int_{-\infty}^{+\infty} e^{-i(\omega_- - \omega_{e''e})\tau_0 x} \tilde{\phi}^*\left(x\right) \tilde{\phi}\left(x+\frac{\Delta T}{\tau_0}\right) dx.
    \end{split}
\label{W}
\end{equation}

Eq.(\ref{W}) indicates a decay of the destructive two-photon interference, when $\Delta T\gg \tau_0$. This evidences a promising HOM effect in the Q-USRS, as being sensitive to a simultaneous arrival of photons \cite{PhysRevLett.61.54}.

We see clearly from Eq.(\ref{SRI}) the time-frequency-resolved nature of the entangled Q-USRS, beyond the spectral and temporal scales conjugated by classical pulses. More advanced information about molecules and environments would be therefore unveiled. These will be elaborated by simulating the Raman signal using certain molecular models.

{\it Microscopic theory}.--We will adopt the molecular aggregate model into the Q-USRS. The molecular Hamiltonian is $H_M=H_0+\sum_{n=1}^N V_n(t)$ with $H_0 = \sum_{n=1}^{N} \omega_n \sigma_n^+ \sigma_n^- - J \sum_{n=1}^{N-1} \left(\sigma_{n+1}^+ \sigma_n^- + \text{h.c.}\right)$ and
\begin{equation}
    \begin{split}
        V_n(t) = \sum_s f_{ns}\sigma_n^+ \sigma_n^- Q_{n,s}(t)
    \end{split}
\label{HM}
\end{equation}
where $Q_{n,s}(t)$ is the coordinate of vibrations and $f_{ns}$ gives the exciton-vibration coupling strength. The model includes $N$ photoactive molecules; $\sigma_n^{+}$ is the raising operator of excitons at the $n$th molecule, in the limit of large onsite exciton-exciton coupling. $\omega_n$ is the exciton energy and $J$ is the hopping rate. A simple model for molecular trimer dyes can thus obtained by $N=3$. Previous studies observed that the absorption/fluorescence spectrum of molecular aggregates show a dense distribution of vibrational states attached to electronic excitations, evident by the inhomogeneous line broadening characterized by a smooth spectral density of vibrations \cite{book1,doi:10.1063/1.460185}.

Averaging over the vibrations and radiative loss, i.e., $\rho(t)=\text{Tr}_B |\psi(t)\rangle\langle \psi(t)|$, the nonradiative relaxation is opened for the Frenkel excitons in the aggregates. We obtain the equation of motion $\dot{\rho}(t) = -i[H_0,\rho(t)] + \hat{\text{{\cal W}}}\rho(t)$ with the Lindbladian $\hat{\text{{\cal W}}}=\sum_{j>i}\hat{\text{{\cal W}}}_{ji}$ that contains the jumping operators between the eigenstates of $H_0$ \cite{SMM}. The density-matrix dynamics therefore provide a microscopic description for the Q-USRS. 


{\it Simulation of Q-USRS}.--We used a Photosystem (PS) trimer model to simulate the USRS with entangled photon pairs combined with nonlinear interferometer scheme. The two-photon wave function in Eq.(\ref{Phi}) is taken to be $A(\omega)=\sigma_0/(\omega^2 + \sigma_0^2)$, $\phi=\frac{2i}{\tau_0}/(\omega_s + \omega_i - \omega_+ + i\frac{2}{\tau_0})$ assuming $\tau_s=\tau_i=\tau_0$ along with before. In what follows, we assert that $\omega_-$ can be tuned whereas fixing $\omega_+$.

\begin{figure}[t]
 \captionsetup{justification=raggedright,singlelinecheck=false}
\centering
\includegraphics[width=0.5\textwidth]{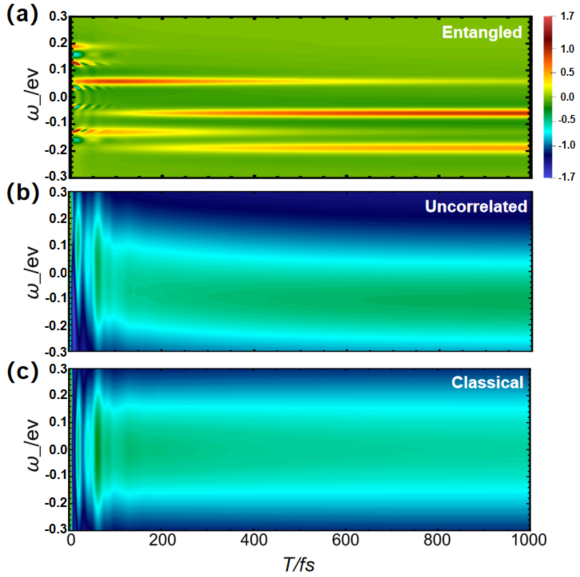}
\caption{(a) Q-FSRS signal with entangled photons from Eq.(\ref{SR}); (b) Q-FSRS signal with uncorrelated photons; (c) FSRS signal using classical pulses. Parameters are: $\sigma_0=1$meV, $\tau_0=25$fs, $\omega_+ = 0.3$eV for (a), and $\tilde{\sigma}_0^{-1} = 2\tau_0 = 50$fs for (b,c); $\omega_1=2.25$eV, $\omega_2=\omega_3=2.1$eV, $J=30$meV from PBI trimers \cite{10.1063/1.2721540}.} 
\label{comparison}
\end{figure}

Fig.\ref{comparison} shows the USRS signal with various photon statistics. The plots vary with $\omega_-$ and arrival time $T$ of the two photons with a zero optical delay ($\Delta T=0$). Using the parameters from PS trimers, we consider $N=3$ molecules and weak coupling $J=30$meV that yields small energy splitting as illustrated in Fig.\ref{set up}(b). 

Fig.\ref{comparison}(a) displays several peaks presenting dramatic different behaviors when varying the delay $T$ (arrival time of entangled photons). In particular, at the Raman shift of $\omega_-=-0.059, -0.189$eV , two long stripe-shaped peaks can be observed promisingly at $T>400$fs. The time-evolving dynamics $\rho_{e_1,e_1}(T+\frac{\tau_0}{2})$ is thus monitored, meaning a downhill transfer of exciton population towards $|e_1\rangle$. Likewise, the peak intensity at $\omega_-=-0.13, 0.059$eV shows a rapid increase during $100$fs and drops smoothly therein. This monitors the dynamics of exciton population at $|e_2\rangle$, i.e.,  $\rho_{e_2,e_2}(T+\frac{\tau_0}{2})$. The two peaks at $\omega_-=0.13, 0.189$eV resolve the exciton dynamics at $|e_3\rangle$, whose population drops dramatically within 700fs. Nevertheless, a few side peaks can be seen within a shorter timescale, by a zoom-in. The oscillatory nature evidences the quantum coherence of excitons, as detailed next.

\begin{figure}[t]
 \captionsetup{justification=raggedright,singlelinecheck=false}
\centering
\includegraphics[scale=0.64]{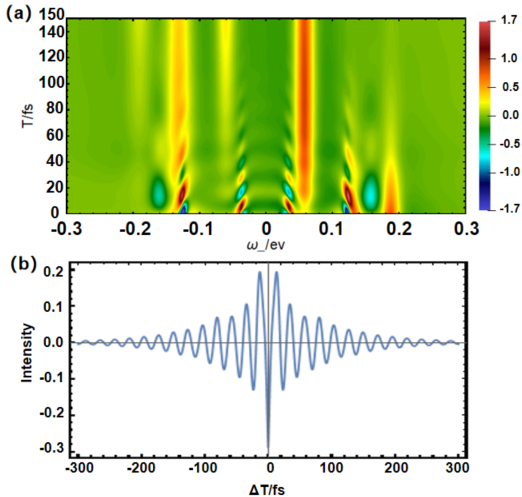}
\caption{(a) Zoom-in of Fig.\ref{comparison}(a) for a short timescale; (b) Q-FSRS signal with entangled photons varies with the optical delay $\Delta T=T_i-T_s$, at the Raman peak $\omega_-=\omega_{e_3,e_2}=0.13$eV. Other parameters are the same as Fig.\ref{comparison}(a).}
\label{Entangled zoom}
\end{figure}

Fig.\ref{Entangled zoom}(a) illustrates the fast dynamics of coherences coexisting with the exciton populations. The side peaks essentially monitor the coherences, provided that $\rho_{ee'} (e\neq e')$ are associated with the Raman resonances distinct from $\rho_{ee}$, as given by Eq.(\ref{SRI}). For instance, the peaks at $\omega_-=\pm 0.124$eV resolve the excitonic coherence $\rho_{e_2,e_3}$ and $\rho_{e_3,e_2}$, when observing an oscillation period of 25fs.

Indeed, the exciton dynamics is subject to a temporal scale $\tau_0$ (pulse duration of an entangled photon pair), whereas the line broadening of the Raman signal is given by $\sigma_0$. These are evident by Eq.(\ref{SRI}) and are confirmed from the simulations. 


Fig.\ref{comparison}(b) and \ref{comparison}(c) plot the USRS without quantum entanglement. The result using uncorrelated twin photons is given in Fig.\ref{comparison}(b), where the each photon has a bandwidth $\tilde{\sigma}_0=\frac{1}{2\tau_0}$. Similar consideration applies to the classical USRS, whereby using two laser pulses with a bandwidth $\tilde{\sigma}_0=\frac{1}{2\tau_0}$ to drive the SRS. Fig.\ref{comparison}(b,c) evidence  
fairly blurred spectral lines and dynamics
if the pulses are short. And the classical USRS dilutes the selectivity of molecular correlation functions, disclosing an intrinsic limit for accessing the excited-state dynamics. This can be seen from the symmetric distribution of the Stokes and anti-Stokes lines in Fig.\ref{comparison}(c). Such a lack of pathway selectivity holds as well for the FSRS using combined narrow- and broad-band pulses.


{\it Multi-photon interference}.--To see closely the  quantum-light-enabled selectivity, we scan the optical delay $\Delta T=T_i - T_s$ so that the s and i photons have different arrival times. Fig.\ref{Entangled zoom}(b) shows a slice of the entangled Q-USRS at $\omega_-=\omega_{e_3,e_2}$, $T=0$. This is a typical HOM interference: a decaying envelop as the two photons get separated in time. The destructive interference is however retained, as a result of light-molecule interactions which yield a residue of the HOM dip at $\Delta T=0$. The dip intensity does not only reveal an overlap between wavepackets of twin photons, but also imprints the electronic structure and dynamics of molecules. The latter is extensively important for spectroscopy and sensing, in that the dip intensity indeed engraves the interference between the two parametric processes given in Fig.\ref{energy level}. The photon pair at outgoing ports interfere with its interchange $s \leftrightarrow i$ essentially.

Moreover, Fig.\ref{Entangled zoom}(b) illustrates a quantum beating with a dominant frequency of $|\omega_+ + \omega_-|/2$ for the anti-Stokes Raman peaks and of $|\omega_+ - |\omega_-||/2$ for Stokes Raman peaks. This makes sense once noting the frequency offset between the two wavepackets of the twin photons \cite{PhysRevLett.61.54}.

{\it Conclusion}.--In summary, (i) we develop a microscopic theory for the USRS with quantum-light fields, incorporating the photon-coincidence counting. (ii) A simple analytic expression is derived for the Q-USRS using entangled photons, yielding elaborate HOM interference which enables the selective access of molecular correlation functions. (iii) Unprecedented time-frequency-resolved property of the Q-USRS is shown explicitly, from the fact that the temporal and spectral scales are not conjugated due to the quantum correlations of photons. (iv) Fast exciton dynamics is visualized including the fluctuating energy-gap and populations in real-time domain. (v) No gratings are needed for spectrally-resolved lines.

All these yield unprecedented scales for spectroscopy to be facilitated with quantum advantage, and would be sufficient for the tomography of density matrix of molecules. Our work, as a new coherent Raman technique, may open a new frontier for studying the ultrafast processes in photochemistry, nano-plasmonics and semiconducting heterostructures.



\begin{acknowledgments}
J.F. and Z.Z. gratefully acknowledge the support of the Early Career Scheme from Hong Kong Research Grants Council (No. 21302721), the National Science Foundation of China (No. 12104380), and the National Science Foundation of China/RGC Collaborative Research Scheme (No. 9054901). Z.Y.O. gratefully acknowledges the support of the General Research Fund from Hong Kong Research Grants Council (No. 11315822).

We would like to thank Kai Wang from the Sun Yat-sen University and Chunfeng Zhang from the Nanjing University for fruitful discussions.
\end{acknowledgments}

\nocite{*}

\providecommand{\noopsort}[1]{}\providecommand{\singleletter}[1]{#1}%

\end{document}